\documentclass[10pt,a4paper]{article}

\usepackage{natbib}
\usepackage{graphicx}
\usepackage{color}
\usepackage{amsmath} 
\usepackage{amssymb} 
\usepackage{bm} 
\usepackage{hyperref}
\usepackage{tikz}
\usepackage{setspace}
\usepackage{multirow}
\usepackage[affil-it]{authblk}

\def\T{{ \mathrm{\scriptscriptstyle T} }}

\newcommand{\Deltab} {\Delta}
\newcommand{\intd} {\textrm{d}}
\newcommand{\Bmat} {B}
\newcommand{\Cmat} {C}
\newcommand{\cmat} {c}
\newcommand{\Imat} {I}

\newcommand{\svec} {s}
\newcommand{\uvec} {u}
\newcommand{\omegab} {\omega}

\newcommand{\s}{s}
\newcommand{\h}{h}

\renewcommand{\v}{v}
\renewcommand{\u}{u}
\newcommand{\w}{w}

\newcommand{\x}{x}
\newcommand{\Y}{Y}
\newcommand{\Yvec}{Y}
\newcommand{\Zvec}{Z}

\newcommand{\bB}{B}
\newcommand{\bbeta}{\beta}
\newcommand{\bzero}{0}
\newcommand{\bSigma}{\Sigma}

\newcommand{\E}{E}
\newcommand{\cov}{\mathrm{cov}}
\newcommand{\var}{\mathrm{var}}

\newcommand{\RR}{\mathbb{R}}

\let\originalleft\left
\let\originalright\right
\renewcommand{\left}{\mathopen{}\mathclose\bgroup\originalleft}
\renewcommand{\right}{\aftergroup\egroup\originalright}

\title{Multivariate Spatial Covariance Models: A Conditional Approach}
\author{Noel Cressie\thanks{\texttt{ncressie@uow.edu.au}}~}
\author{Andrew Zammit-Mangion\thanks{\tt{azm@uow.edu.au}}}
\affil{National Institute for Applied Statistics Research Australia~(NIASRA),\\ School of Mathematics and Applied Statistics,\\ University of Wollongong, New South Wales 2522, Australia}
\date{}

\makeatletter
\renewcommand\section{\@startsection{section}{1}{\z@}%
                                  {-3.5ex \@plus -1ex \@minus -.2ex}%
                                  {2.3ex \@plus.2ex}%
                                  {\normalfont\large\bfseries}}
\makeatother

\begin{document}

\maketitle

\begin{abstract}
Multivariate geostatistics is based on modelling all covariances between all possible combinations of two or more variables at any sets of locations in a continuously indexed domain. Multivariate spatial covariance models need to be built with care, since any covariance matrix that is derived from such a model must be nonnegative-definite. In this article, we develop a conditional approach for spatial-model construction whose validity conditions are easy to check. We start with bivariate spatial covariance models and go on to demonstrate the approach's connection to multivariate models defined by networks of spatial variables. In some circumstances, such as modelling respiratory illness conditional on air pollution, the direction of conditional dependence is clear. When it is not, the two directional models can be compared. More generally, the graph structure of the network reduces the number of possible models to compare. Model selection then amounts to finding possible causative links in the network. We demonstrate our conditional approach on  surface temperature and pressure data, where the role of the two variables is seen to be asymmetric.
\end{abstract}

\section{Introduction}\label{sec:Intro}

The conditional approach to building multivariate spatial covariance models was introduced by \citet{Royleetal1999}. 
In that article, pressure and wind fields are modelled as a bivariate process over a region of the globe, with the wind process conditioned on the pressure process through a physically-motivated stochastic partial differential equation. 
In general, such models exhibit asymmetry; that is, for $Y_1(\cdot)$ and $Y_2(\cdot)$ defined on $d$-dimensional Euclidean space $\mathbb{R}^d$,
\begin{equation*}
\cov\left\{Y_1(\s),Y_2(\u)\right\}\neq \cov\left\{Y_2(\s),Y_1(\u)\right\},\quad \svec,\uvec \in \mathbb{R}^d.
\end{equation*}
Of course, it is always true that  $\cov\left\{Y_1(\svec),Y_2(\uvec)\right\} = \cov\left\{Y_2(\uvec),Y_1(\svec)\right\}$.

There are commonly-used classes of multivariate spatial models that assume symmetric, stationary dependence in the cross-covariances; that is, they assume 
$C_{12}(\h)\equiv\cov\left\{Y_1(\s),Y_2(\s+\h)\right\}=\cov\left\{Y_2(\s),Y_1(\s+\h)\right\}\equiv C_{21}(\h)$, for $\h \in \mathbb{R}^d$ (e.g., \citealp{Gelfandetal2004};~\citealp[Section 4.1.5]{CressieWikle2011};~\citealp{GentonKleiber2015}). The most notable of these symmetric-cross-covariance models is the linear model of coregionalization; see, for example, \citet[][Section III.B.3]{JournelHuijbregts1978}, \citet{Websteretal1994}, \citet{Wackernagel1995},~and~\citet[Section~9.5]{Banerjeeetal2015}. While symmetry may reduce the number of parameters or allow fast computations, it may not be supported by the underlying science or by the data. 

\citet{VerHoefCressie1993} avoid making symmetry restrictions by working with variance-based cross-variograms and propose a spatial shift parameter to express asymmetry. \citet{GentonKleiber2015} review other approaches that capture asymmetry and include those of \citet{Apanasovich_2010} and \citet{Li_2011}; see also \citet{ChristensenAmemiya2001}. In multivariate spatial-lattice modelling, \citet{SainCressie2007}, \citet{Sainetal2011}, and \citet{Martinez_2013} specifically include asymmetry in their models. 

A key outcome of multivariate geostatistics is optimal spatial prediction of a hidden multivariate spatial process, $\Y(\cdot)=\left\{Y_1(\cdot),\ldots,Y_p(\cdot)\right\}^\T$, based on multivariate noisy spatial observations, $\{Z_q(\s_{qi}):i=1,\ldots,m_q,\,q=1,\ldots,p\}$, of the hidden processes $\{Y_q(\cdot): q = 1,\dots,p\}$. Assuming additive measurement error, $\varepsilon_{q}(\cdot)$, we have data $Z_q(\cdot)=Y_q(\cdot)+\varepsilon_q(\cdot)$ at the $m_q$ data locations, $D_q^O \equiv \{\s_{qi}:i=1,\ldots,m_q\}$, for $q=1,\ldots,p$. Notice that we have not assumed colocated data for the different spatial variables. Optimally predicting just one of the processes, say $Y_1(\cdot)$, using the multivariate data $\{Z_q(\svec_{qi})\}$, is often called cokriging. 

Contributions to multivariate-spatial-prediction methodology include those of 
\citet{Myers1982, Myers1992}, 
\citet{VerHoefCressie1993}, 
\citet{Wackernagel1995}, 
\citet{CressieWikle1998b}, 
\citet{RoyleBerliner1999}, 
\citet{Gelfandetal2004}, 
\citet{MajumdarGelfand2007}, 
\citet{Finleyetal2008}, 
\citet{Huangetal2009}, 
\citet[Section~4.1.5]{CressieWikle2011}, 
\citet{Furrer_2011},
\citet{Heaton_2011}, and 
\citet[Chapter 7]{Banerjeeetal2015}.


\citet{GentonKleiber2015} give a comprehensive review of many different ways that valid multivariate covariances can be constructed, with a brief mention of the conditional approach. 
For spatial-lattice data, \citet{Kim_2001} and~\citet{Jin_2005} use a conditional approach to modelling multivariate spatial dependence. For regularly or irregularly gridded spatial processes, \citet[p.~234]{CressieWikle2011} clarify the discussion of the conditional approach given in \citet{Gelfandetal2004}. For geostatistical data, \citet{Heaton_2011} build a multivariate model for predicting $Z_2$ from $Z_1$ by conditioning on $Z_1$ and a kernel-smoothed $\widetilde{Z}_1$. In this article, we show that \citet{Royleetal1999} and \citet{Heaton_2011} describe specific cases of a large class of multivariate models whose existence we establish.



\section{Modelling joint dependence through conditioning}\label{sec:2}

In this section, we introduce the conditional approach by considering the bivariate case. Here, $\{(Y_1(\s),Y_2(\s)):\s\in D\subset \RR^d\}$ are two co-varying spatial processes in a continuous-spatially-indexed domain $D$ of positive volume contained in $d$-dimensional Euclidean space $\mathbb{R}^d$; the multivariate case is considered in Section~\ref{sec:4}. As was seen in Section \ref{sec:Intro}, it is sometimes convenient to write the individual processes as $Y_1(\cdot)$ and $Y_2(\cdot)$, respectively. Then the joint probability measure of $Y_1(\cdot)$ and $Y_2(\cdot)$ can be written as,
\begin{equation}\label{eqn:joint-prob-measure}
[Y_1(\cdot),Y_2(\cdot)]=[Y_2(\cdot)\mid Y_1(\cdot)][Y_1(\cdot)],
\end{equation}
where we use the convention that $[A\mid B]$ represents the conditional probability of $A$ given $B$, and $[B]$ represents the marginal probability of $B$. The conditional probability in \eqref{eqn:joint-prob-measure} is shorthand for $[\{Y_2(\svec):\svec\in D\}|\{Y_1(v):v\in D\}]$, which we see in Section \ref{sec:3} is defined through the finite-dimensional distributions.
In this article, we are particularly interested in the conditional distributional properties of $Y_2(\s)$ and of $\left\{Y_2(\s),Y_2(\u)\right\}$, given $\{Y_1(v): v \in D\}$.

The order of the variables is a choice, but it is generally driven by the underlying science; for example, $Y_1(\cdot)$ might be ambient ozone in a city and $Y_2(\cdot)$ might represent the spatial intensity or density of respiratory illness in the city; or $Y_1(\cdot)$ might be a temperature field and $Y_2(\cdot)$ might be a rainfall field, where $Y_2(\cdot)$ depends to some extent on $Y_1(\cdot)$ through evapo-transpiration and the Penman--Monteith equation \citep[e.g.,][]{Beven1979}. When the order is not obvious, both models can be fitted and the best one selected, indicating discovery of a possible causative link. For the multivariate case in Section \ref{sec:4}, it is enough to have a partial order on the variables or, equivalently, a directed acyclic graph \citep{Cressie_1998}.

Assume that $\E\{Y_1(\cdot)\}\equiv 0 \equiv \E\{Y_2(\cdot)\}$; we relax this in Section~\ref{sec:3}. Consider the following model for the first two conditional moments of $[\{Y_2(\svec):\svec\in D\}\mid Y_1(\cdot)]$:
\begin{align}\label{eqn:E-and-cov}
\E\left\{Y_2(\s)\mid Y_1(\cdot)\right\}&\,=\int_D{b(\s,\v)Y_1(\v)\,\intd \v},\quad \s\in D,\\
\cov\left\{Y_2(\s),Y_2(\u)\mid Y_1(\cdot)\right\}&\,=C_{2{\mid} 1}(\s,\u),\quad \s,\u\in D,\label{eqn:E-and-cov2}
\end{align}
where $b(\cdot,\cdot)$ is any integrable function that maps from $\RR^d\times \RR^d$ into $\RR$, and $C_{2\mid 1}(\cdot,\cdot)$ is a univariate covariance function that does not depend functionally on $Y_1(\cdot)$. In \eqref{eqn:E-and-cov}, $b(\cdot,\cdot)$ may be obtained from scientific understanding of how $Y_2(\cdot)$ evolves from $\{Y_1(\v) : \v \in D\}$. Hence, we call $b$ an interaction function, and it has an important role in scientific modelling of positive or negative dependence of $Y_2$ on $Y_1$. Recall from Section \ref{sec:Intro} that $Y_q$ is observed with measurement error, resulting in $Z_q$, for $q = 1,2$. Unlike in \citet{Royleetal1999} and \citet{Heaton_2011}, the focus of \eqref{eqn:E-and-cov} and \eqref{eqn:E-and-cov2} is on the latent processes $Y_1$ and $Y_2$, rather than on $Z_1$ and $Z_2$. Important special cases of \eqref{eqn:E-and-cov} include $b(\s,\v)$ proportional to a kernel smoothing function and $b(\s,\v)$ proportional to a Dirac delta function, which describes pointwise dependence.


Critically, the conditional covariance function $C_{2\mid 1}$ in \eqref{eqn:E-and-cov} is necessarily a nonnegative-definite function, and there are many classes of such functions available (e.g.,~\citealp{Christakos1984}; \citealp[Section~2.5]{Cressie1993}; \citealp[Section~2.2]{Banerjeeetal2004}). Finally, suppose that $Y_1(\cdot)$ has a valid univariate covariance function $C_{11}(\cdot,\cdot)$, which is also necessarily nonnegative-definite. Thus, the conditional approach requires only specification of an integrable interaction function and two valid univariate spatial covariance functions, $C_{2\mid 1}$ and $C_{11}$, leading to rich classes of cross-covariance functions. Section \ref{sec:cross-cov} gives one such class. 

Define $C_{qr}(\s,\u)\equiv \cov\left\{Y_{q}(\s),Y_r(\u)\right\}$, for $q,r=1,2$ and $\s,\u\in D$. From the two univariate spatial covariance models, $C_{2\mid 1}$ and $C_{11}$, we have:
\begin{align}
C_{22}(\s,\u)&\equiv\cov\left\{Y_2(\s),Y_2(\u)\right\} \nonumber \\
&=\cov\left[\E\left\{Y_2(\s)\mid Y_1(\cdot)\right\},\E\left\{Y_2(\u)\mid Y_1(\cdot)\right\}\right] + E\left[\cov\left\{Y_2(\s),Y_2(\u)\mid Y_1(\cdot)\right\}\right] \nonumber\\
&=\int_D\int_{D}{b(\s,\v)C_{11}(\v,\w)b(\u,\w)\,\intd\v\intd\w} + \,C_{2\mid 1}(\s,\u),\quad\s,\u\in D.\label{eqn:cov1}
\end{align}

When $\uvec=\svec$ in~\eqref{eqn:cov1}, one can see that $\var\{Z_2(\svec)\}$ can be expressed as a decomposition of spatial variation due to its regression on $Y_1(\cdot)$ plus the remaining variation, $C_{2\mid 1}(\svec,\svec)$, unexplained by $Y_2$'s dependence on $Y_1$. In general,~\eqref{eqn:cov1} shows a decomposition of spatial covariation into an explanatory component and a descriptive component.

Importantly, the formulas for the cross-covariances are straightforward:
\begin{equation}
C_{12}(\s,\u)=\,\cov\left[Y_1(\s),E\{Y_2(\u)\mid Y_1(\cdot)\}\right]=\,\int_D{C_{11}(\s,\w)b(\u,\w)\,\intd \w},\quad\s,\u\in D,\label{eqn:cov2}
\end{equation}
which has only an explanatory component. The other cross-covariance is obtained from
\begin{equation}\label{eqn:C21}
 C_{21}(\s,\u)=C_{12}(\u,\s),\quad \s,\u\in D.
\end{equation}
Finally, recall that
\begin{equation}\label{eqn:C11}
C_{11}(\s,\u)=\cov\left\{Y_1(\s),Y_1(\u)\right\},\quad\s,\u\in D,
\end{equation}
is a given nonnegative-definite function, and this is descriptive only of spatial covariation in $Y_1$. 

Then \eqref{eqn:cov1}--\eqref{eqn:C11} specifies all covariances $\{C_{qr}(\cdot,\cdot)\}$, and any covariance matrix obtained from them will be nonnegative-definite; see Section~\ref{sec:3}. From \eqref{eqn:cov2}, $C_{12}(\uvec,\svec) = \int_DC_{11}(\uvec,\w)b(\svec,\w)\intd\w \ne C_{12}(\svec,\uvec)$, in general, because $b(\cdot,\cdot)$ may be asymmetric. That is, the conditional approach captures asymmetry naturally through the interaction function.


\section{Bivariate stochastic processes based on conditioning}\label{sec:3}


\subsection{Existence of a bivariate stochastic process}\label{sec:3-1}

Let $[\{Y_1^0(\s),Y_2^0(\s)\}:\s\in \mathbb{R}^d]$ be a bivariate Gaussian process with mean $\bzero$, covariance functions $C_{11}^0(\cdot,\cdot)$, $C_{22}^0(\cdot,\cdot)$, and cross-covariance functions $C_{12}^0(\cdot,\cdot), C_{21}^0(\cdot,\cdot)$. Then for any pair of nonnegative integers $n_1,n_2$ such that $n_1 + n_2 > 0$; for any locations $\{\s_{1k}:k=1,\ldots,n_1\}$, $\{\s_{2l}:l=1,\ldots,n_2\}$, and for any real numbers $\{a_{1k}:k=1,\ldots,n_1\}$, $\{a_{2l}:l=1,\ldots,n_2\}$,
\begin{align}
  &\var\left\{\sum_{k=1}^{n_1}{a_{1k}Y_1^0(\s_{1k})}+\sum_{l=1}^{n_2}{a_{2l}Y_2^0(\s_{2l})}\right\}\nonumber \\
  &\quad=\sum_{k=1}^{n_1}\sum_{k'=1}^{n_1} a_{1k}a_{1k'}C_{11}^0(\s_{1k},\s_{1k'})+\sum_{l=1}^{n_2}\sum_{l'=1}^{n_2} a_{2l}a_{2l'}C_{22}^0(\s_{2l},\s_{2l'}) \nonumber\\
  &\qquad+\sum_{k=1}^{n_1}\sum_{l'=1}^{n_2} a_{1k}a_{2l'}C_{12}^0(\s_{1k},\s_{2l'})+\sum_{l=1}^{n_2}\sum_{k'=1}^{n_1} a_{2l}a_{1k'}C_{21}^0(\s_{2l},\s_{1k'})~ \ge 0.\label{eqn:var-ge-zero}
\end{align}

Conversely, suppose that the set of functions, $\{C_{qr}(\cdot,\cdot):q,r=1,2\}$, has the property that 
$C_{12}(\s,\u)=C_{21}(\u,\s)$, for all $\s,\u\in \mathbb{R}^d$, and that \eqref{eqn:var-ge-zero} holds. Then there exists a bivariate Gaussian process $\{(Y_1(\s),Y_2(\s)):\s\in \RR^d\}$ such that
\begin{equation*}
\cov\{Y_q(\s),Y_r(\u)\}= C_{qr}(\s,\u),\quad \s,\u\in \RR^d; q,r=1,2.
\end{equation*}
The proof of this result relies on establishing the Kolomogorov consistency conditions \citep[e.g.,][pp.~482--484]{Billingsley1995} for the finite-dimensional distributions of
\begin{equation*}
\{Y_1(\s_{11}),\ldots,Y_1(\s_{1n_1}),Y_2(\s_{21}),\ldots,Y_2(\s_{2n_2})\}.
\end{equation*}
They are specified to be Gaussian with second-order moments defined by \eqref{eqn:cov1}--\eqref{eqn:C11}. The consistency conditions are: the finite-dimensional distributions are consistent over marginalization; and permutation of the variables' indices does not change the probabilities of events, which we now establish.

Consider $\{C_{qr}(\cdot,\cdot)\}$ defined by \eqref{eqn:cov1}--\eqref{eqn:C11}. Because the finite-dimensional distributions are Gaussian, permutation-invariance is guaranteed by \eqref{eqn:C21}, an expression for covariances.
The right-hand side of \eqref{eqn:cov1} consists of $C_{2\mid 1}(\cdot,\cdot)$, which is nonnegative-definite, added to a quadratic term that is guaranteed to be nonnegative-definite, since $C_{11}(\cdot,\cdot)$ in \eqref{eqn:C11} is nonnegative-definite. Hence, $C_{22}(\cdot,\cdot)$, which is the sum of these two terms, is nonnegative-definite. Thus, marginally, $Y_2(\cdot)$ has a nonnegative-definite covariance function, but this is not enough. It remains to establish~\eqref{eqn:var-ge-zero}. Substitute \eqref{eqn:cov1}~and~\eqref{eqn:cov2} into the left-hand side of \eqref{eqn:var-ge-zero} to obtain 
\begin{equation}\label{eqn:left-hand-of-var}
\sum_{l=1}^{n_2}\sum_{l'=1}^{n_2}a_{2l}a_{2l'}C_{2\mid 1}(\s_{2l},\s_{2l'})+\int_D \int_D{a(\s)a(\u)C_{11}(\s,\u)\,\intd\s\intd\u},
\end{equation}
where for $\delta(\cdot)$ the Dirac delta function,
\begin{equation*}
a(\s)\equiv \sum_{k=1}^{n_1}a_{1k}\delta(\s-\s_{1k})+\sum_{l=1}^{n_2}a_{2l}b(\s_{2l},\s),\quad \s\in \RR^d.
\end{equation*}
Since $C_{2\mid 1}$ and $C_{11}$ are nonnegative-definite, \eqref{eqn:left-hand-of-var} is nonnegative, resulting in \eqref{eqn:var-ge-zero}.

Only nonnegative-definite functions for univariate processes are needed in the conditional approach. Further, the finite-dimensional distribution, \begin{align*}&[\{Y_1(\svec_{1k}),Y_2(\svec_{2l}):k=1,\dots,n_1;\, l = 1,\dots,n_2\}] \\&~~~= [\{Y_2(\svec_{2l}):l=1,\dots,n_2\} \mid \{Y_1(\svec_{1k}):k=1,\dots,n_1\}][\{Y_1(\svec_{1k}):k=1,\dots,n_1\}],\end{align*} depends critically on the finite collection of interaction functions, $\{b(\s_{2l},\cdot):l=1,\ldots,n_2\}$. The only condition we place on $b(\cdot,\cdot)$ is that it is a real-valued integrable function.

The existence proof given above shows that there is at least one process with covariance functions given by~\eqref{eqn:cov1}--\eqref{eqn:C11}. However, the modeller is not restricted to fitting bivariate Gaussian processes.~\cite{Zammit_2016} fit a non-Gaussian model constructively through \eqref{eqn:E-and-cov}.

In practice, geostatistical software will discretize the continuous spatial domain $D$ onto a fine-resolution finite grid defined by the spatial lattice, $D^L \equiv\{\s_1,\ldots,\s_n\}$, which represents the centroids of the grid cells. That is, $Y_q(\cdot)$ is replaced with the vector $\Y_q\equiv \left\{Y_q(\s_1),\ldots,Y_q(\s_n)\right\}^\T,~q=1,2$. Under this discretization, \eqref{eqn:cov1}--\eqref{eqn:C11} become, respectively, 
\begin{align}
\cov(\Y_2)=&\,\bSigma_{2\mid 1}+\bB\bSigma_{11}\bB^\T,\label{eqn:cov-Y2}\\
\cov(\Y_1,\Y_2)=&\,\bSigma_{11}\bB^\T,\label{eqn:cov-Y1-Y2}\\
\cov(\Y_2,\Y_1)=&\,\bB\bSigma_{11},\label{eqn:cov-Y2-Y1}\\
\cov(\Y_1)=&\,\bSigma_{11},\label{eqn:cov-Y1}
\end{align}
which were given by \citet[p.~160]{CressieWikle2011} and were used by \citet{Jin_2005} for modelling bivariate spatial-lattice data. In~\eqref{eqn:cov-Y2}--\eqref{eqn:cov-Y1}, $\bSigma_{2\mid 1}$~and~$\bSigma_{11}$ are nonnegative-definite $n\times n$ covariance matrices obtained from $\{C_{2\mid 1}(\s_k,\s_l):k,l=1,\ldots,n\}$ and $\{C_{11}(\s_k,\s_l):k,l=1,\ldots,n\}$, respectively, and $\bB$ is the square $n\times n$ matrix obtained from $\{b(\s_k,\s_l):k,l=1,\ldots,n\}$. Hence, the following $2n\times 2n$ joint covariance matrix is nonnegative-definite:
\begin{equation}\label{eqn:cov-matrix}
\cov\left\{ (\Yvec_1^\T, \Yvec_2^\T)^\T\right\} =  \begin{bmatrix}\bSigma_{11} & \bSigma_{11}\bB^\T \\ \bB \bSigma_{11} & \bSigma_{2\mid 1}+\bB\bSigma_{11}\bB^\T  \end{bmatrix}.
\end{equation}

\citet[p.~273]{Banerjeeetal2015} state that it is meaningless to talk about the joint distribution of $Y_2(\s_1)\mid Y_1(\s_1)$ and $Y_2(\s_2)\mid Y_1(\s_2)$ as building blocks for the conditional approach, with which we agree. They also go on to say that this ``reveals the impossibility of conditioning,'' with which we disagree. We have shown in this section that the conditional approach yields a well-defined bivariate Gaussian process $\{Y_1(\cdot),Y_2(\cdot)\}$, since conditioning is on the whole process $Y_1(\cdot)$. This implies a well-defined joint distribution of the random vectors $\Y_1$ and $\Y_2$, obtained from discretization, whose joint distribution is given by $[\Y_1,\Y_2]=\, [\Y_2\mid \Y_1][\Y_1]$, where $[\Y_2\mid \Y_1]$ is a $N(\bB\Y_1,\bSigma_{2\mid 1})$ density, and $[\Y_1]$ is a $N(\bzero,\bSigma_{11})$ density. This relation is deceptively simple, but the existence proof above shows how such relations are founded in the joint probability measure of $Y_1(\cdot)$ and $Y_2(\cdot)$.

The conditional density $[Y_2 \mid Y_1]$ is derived from a linear regression of $Y_2$ on $Y_1$, not on the observed variable $\Zvec_1$. The errors-in-variable model (\citealp{Berkson_1950}; \citealp{Heaton_2011}) considers a regression of noisy observations $\Zvec_2$ on noisy observations $\Zvec_1$, which is different from the approach we take. For our conditional approach, the conditioning is on the whole vector $\Y_1$, but any marginal or conditional finite-dimensional distribution can be easily derived. For example, $[Y_2(\s_1)\mid Y_1(\s_1)]$ can be obtained from $[Y_1(\svec_1),Y_2(\svec_1)]/[Y_1(\svec_1)]$, as follows. The numerator is
\begin{equation*}
  [Y_1(\s_1),Y_2(\s_1)]=\int_{\mathbb{R}}\cdots\int_{\mathbb{R}}[Y_2(\s_1)\mid \Y_1][\Y_1]\intd Y_1(\s_2)\ldots\intd Y_1(\s_n),
\end{equation*}
which from \eqref{eqn:cov-matrix} is Gaussian with mean $\bzero$ and $2\times 2$ covariance matrix,
\begin{equation*}
\begin{bmatrix} C_{11}(\s_1,\s_1) & \sum_{k=1}^{n}C_{11}(\s_1,\s_k)b_{1k} \\ \sum_{k=1}^{n}C_{11}(\s_1,\s_k)b_{1k} & ~~~~~C_{2\mid 1}(\s_1,\s_1)+\sum_{k=1}^n\sum_{l=1}^n b_{1k}C_{11}(\s_k,\s_l)b_{1l} \end{bmatrix},
\end{equation*}
\noindent where $b_{ik}$ is the $(i,k)$th element of $\Bmat$ in \eqref{eqn:cov-Y2}--\eqref{eqn:cov-Y2-Y1}, and the denominator is $N(0,C_{11}(\svec_1,\svec_1))$.

We have seen above that it is not just one or a few finite-dimensional distributions that define the conditional approach, it is all of them. \citet[p.~273]{Banerjeeetal2015} state that the conditional approach is flawed and that kriging is not possible. In Section \ref{sec:3-2}, we give a simple, one-dimensional example of the conditional approach defined by \eqref{eqn:cov1}--\eqref{eqn:C11} with kriging and cokriging equations for predicting $\{Y_1(\s_0): \s_0 \in D^L\}$ from noisy incomplete data, $\{Z_q(\svec_{qi}):i=1,\dots,m_q,\, q=1,2 \}$. We deliberately chose not to predict the dependent variable $Y_2$ to illustrate the flexibility of having a fully bivariate model. \cite{Zammit_2015a} show that the important scientific problem of predicting methane fluxes results in cokriging of this type.

The incorporation of non-zero mean functions in $\left\{Y_1(\cdot),Y_2(\cdot)\right\}$ is straightforward. Let $\mu_1(\cdot)$ and $\mu_2(\cdot)$ be real-valued functions defined on $\RR^d$, and suppose that the finite-dimensional Gaussian distributions obtained from $\{Y_1(\s_{1k}), Y_2(\s_{2l}) : k=1,\ldots,n_1;\,l=1,\dots,n_2\}$ have means $\{\mu_1(\s_{1k}),\mu_2(\s_{2l}) : k=1,\ldots,n_1;\,l=1,\ldots,n_2\}$, respectively. Then the method of proof at the beginning of this section yields a bivariate Gaussian process $\left\{Y_1(\cdot),Y_2(\cdot)\right\}$ with mean functions $\{\mu_1(\cdot),\mu_2(\cdot)\}$ and covariance functions $\{C_{qr}(\cdot,\cdot):q,r=1,2\}$. Covariates $\x_1(\cdot)$ and $\x_2(\cdot)$ can then be incorporated through $\mu_q(\s)=\x_q(\s)^\T\bbeta_q~(\s\in D,~q=1,2)$, where $\bbeta_1$ and $\bbeta_2$ are vectors of regression coefficients of possibly different dimensions.

\subsection{Cokriging using covariances defined by the conditional approach}\label{sec:3-2}

Section \ref{sec:3-1} establishes the existence of the bivariate process $\{Y_1(\cdot),Y_2(\cdot)\}$ with $\{C_{qr}(\cdot,\cdot)\}$ given by \eqref{eqn:cov1}--\eqref{eqn:C11}, and hence we may use cokriging for multivariate spatial prediction in the presence of incomplete, noisy data.

The aim of cokriging is to predict, say, $Y_1(\svec_0),~\svec_0 \in D$, based on data $\Zvec_1$ and $\Zvec_2$ \citep[e.g.,][p.~138]{Cressie1993}, where
\begin{equation}\label{eq:DO}
\Zvec_q \equiv \{Z_q(\svec):\svec \in D_q^O\}^\T,\quad D_q^O \equiv \{\svec_{qi}: i=1,\dots,m_q\},~ q=1,2.
\end{equation}
Recall that 
$Z_q(\svec_{qi}) = Y_q(\svec_{qi}) + \varepsilon_q(\svec_{qi})$, 
$\E\{\varepsilon_q(\cdot)\} = 0$, and 
$\var\{\varepsilon_q(\cdot)\} = \sigma^2_{\varepsilon_q}$ $(i = 1,\dots,m_q;\,q=1,2)$. Then, the best predictor for $Y_1(\svec_0)$ is the conditional mean, $\E\{Y_1(\svec_0) \mid  \Zvec_1, \Zvec_2\}$. Assuming mean-zero joint Gaussian processes,
\begin{equation}\label{eq:cokrig}
\hat Y_1(\svec_0) \equiv \E\{Y_1(\svec_0) \mid  \Zvec_1, \Zvec_2\} = \begin{bmatrix} \cmat_{11}^\T & \cmat_{12}^\T \end{bmatrix} 
										\begin{bmatrix} \Cmat_{11} + \sigma^2_{\varepsilon_1} \Imat_{m_1} & \Cmat_{12} \\ \Cmat_{21} & \Cmat_{22} + \sigma^2_{\varepsilon_2} \Imat_{m_2} \end{bmatrix}^{-1}
										\begin{bmatrix} \Zvec_1 \\ \Zvec_2 \end{bmatrix},
\end{equation}
\noindent where for $q,r = 1,2$,
\begin{equation*}
\cmat_{1r}^\T \equiv \{C_{1r}(\svec_0,\svec_{ri}) : i = 1,\dots,m_r \},\,\, \Cmat_{qr}  \equiv \{C_{qr}(\svec_{qi},\svec_{rj}) : i = 1,\dots,m_q,\, j = 1,\dots,m_{r}\},
\end{equation*}
\noindent and $\Imat_{m_q}$ is the $m_q \times m_q$ identity matrix. Expression (\ref{eq:cokrig}) is called the simple-cokriging predictor, and it is also the best linear predictor of $Y_1(\svec_0)$.

While in some multivariate models, the matrices $(\Cmat_{qr}: q,r = 1,2)$ are known in closed form \citep{GentonKleiber2015}, this is not necessarily so here. Cokriging using the conditional approach may require integrations over $D$ in order to compute $(\Cmat_{qr})$. There are examples where the integrals can be carried out analytically. One such example is given in Appendix 1.

\begin{figure}[!t]
\includegraphics[width=\textwidth]{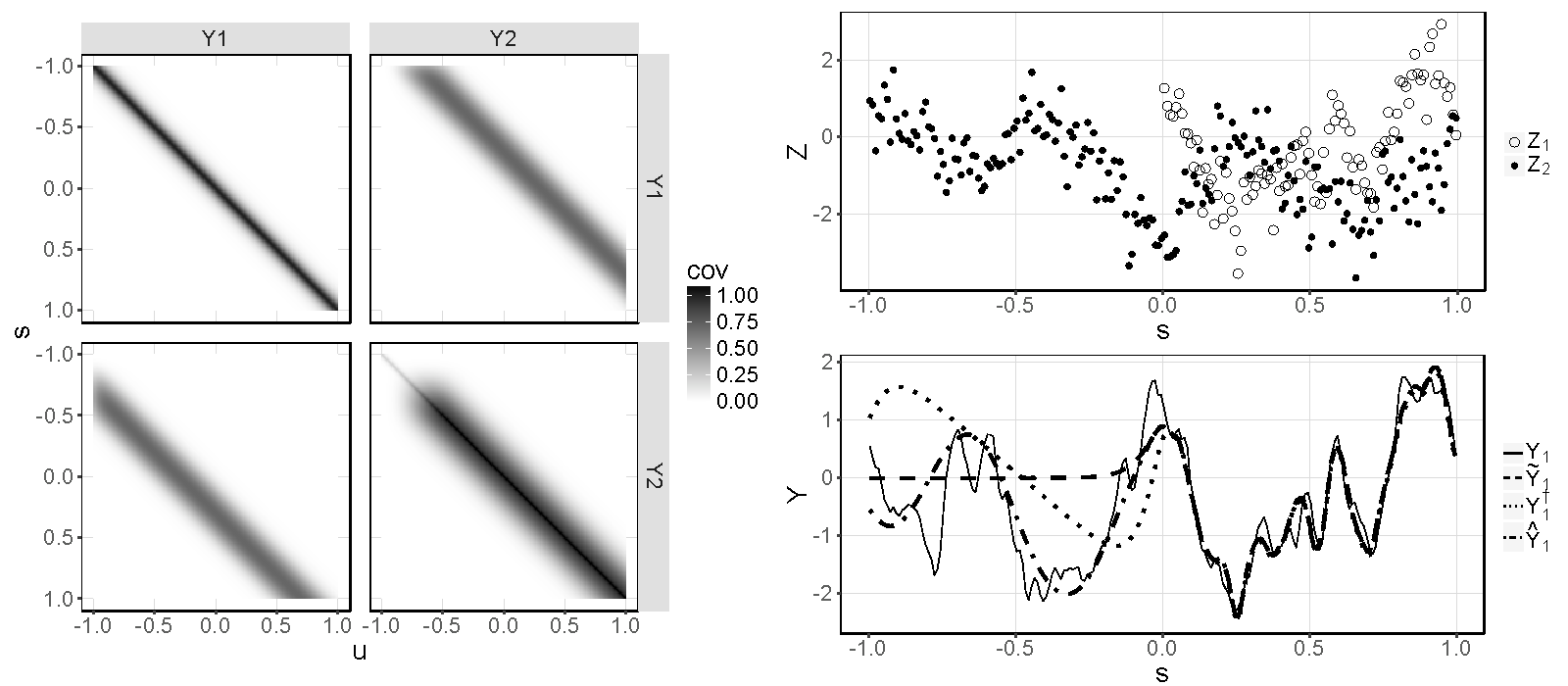}
	\caption{Cokriging using spatial covariances defined by the conditional approach. Left panel: The covariance matrix \eqref{eqn:cov-matrix}. Right panel, top: The simulated observations $\Zvec_1$ (open circles) and $\Zvec_2$ (dots). Right panel, bottom: The hidden value $\Yvec_1$ (solid line), the kriging predictor $\widetilde\Yvec_1$ (dashed line), the misspecified cokriging predictor $\Yvec_1^\dagger$ (dotted line), and the cokriging predictor $\hat\Yvec_1$ (dotted-dashed line).}\label{fig:sim}
\end{figure}

To demonstrate the benefits of cokriging based on a bivariate spatial model defined by the conditional approach, we simulated the processes $Y_1, Y_2$ on a regular discretisation, $D^L$, of $D = [-1,1]$, where $|D^L| = 200$. We describe the covariations in $C_{11}(\cdot,\cdot)$ and $C_{2\mid 1}(\cdot,\cdot)$ by Mat{\'e}rn covariance functions,
 \begin{align}\label{eq:Matern1}
 C_{11}(s,u) &\equiv \frac{\sigma^2_{11}}{2^{\nu_{11}-1}\Gamma(\nu_{11})}(\kappa_{11} |u-s|)^{\nu_{11}}K_{\nu_{11}}(\kappa_{11} |u-s|),\\
 C_{2\mid 1}(s,u) &\equiv \frac{\sigma^2_{2\mid 1}}{2^{\nu_{2\mid 1}-1}\Gamma(\nu_{2\mid 1})}(\kappa_{2\mid 1} |u-s|)^{\nu_{2\mid 1}}K_{\nu_{2\mid 1}}(\kappa_{2\mid 1} |u-s|), \label{eq:Matern2}
 \end{align}
where we set the variances to $\sigma^2_{11}=1, \sigma^2_{2\mid 1}=0.2$, the scale parameters to $\kappa_{11}=25, \kappa_{2\mid 1}=75$, the smoothness parameters to $\nu_{11}=\nu_{2\mid 1}=1.5$, and where $K_\nu$ is a Bessel function of the second kind of order $\nu$. For the interaction function, we used the shifted bisquare function
 \begin{equation}\label{eq:bisquare1D}
 b(s,v) \equiv \left\{\begin{array}{ll} A\{1 - (|v- s - \Delta|/r)^2\}^2, &| v -s  - \Delta| \le r, \\ 
 0, & \textrm{otherwise}, \end{array} \right. 
 \end{equation}
 \noindent where we set the shift parameter to $\Delta = -0.3$ to capture asymmetry, we set the aperture parameter to $r = 0.3$, and we set the scaling parameter to $A = 5$.  The grid cells were used to define the discretized domain over which we carried out the numerical integrations in \eqref{eqn:cov1} and \eqref{eqn:cov2}. For example, $C_{12}(s_0,u) \simeq \sum_{k=1}^{n} \eta_k C_{11}(s_0,w_k)b(u,w_k),$ where $D^L \equiv (w_k : k=1,\dots,n)$ and $(\eta_k: k = 1,\dots,n)$ are the grid spacings. Here $\eta_1 = \cdots = \eta_{200} = 0.01$.  The covariance matrix \eqref{eqn:cov-matrix} is shown in Fig.~\ref{fig:sim}, left panel, where asymmetry is clearly present. Finally, the data $\Zvec_1$ and $\Zvec_2$ in \eqref{eq:DO} were generated by adding independent, mean-zero Gaussian measurement errors with variances $\sigma^2_{\epsilon_1} = \sigma^2_{\epsilon_2} = 0.25$ to $\Yvec_1$ and $\Yvec_2$ at given locations $D_1^O$ and $D_2^O$, respectively. Here, $D_2^O \equiv D^L$ and $D_1^O \equiv D^L \cap [0,1]$, so that $\Yvec_1$ is observed only for $s \ge 0$.

We used the cokriging equation \eqref{eq:cokrig} to obtain $\hat\Yvec_1 \equiv \{\hat Y_1(\svec_0):\svec_0 \in D^L\}^\T$ based on the simulated observations $\Zvec_1$ and $\Zvec_2$. We compared $\hat\Yvec_1$ to a kriging predictor $\widetilde \Yvec_1$ based only on data $\Zvec_1$, where $\widetilde\Yvec_1 \equiv \{\widetilde Y_1(\svec_0) : \svec_0 \in D^L\}^\T$ and $\widetilde Y_1(\svec_0) \equiv \cmat_{11}^\T(\Cmat_{11} + \sigma^2_{\varepsilon_1}\Imat_{m_1})^{-1}\Zvec_1$. We also compared $\widetilde\Yvec_1$ to a misspecified cokriging predictor $\Yvec_1^\dagger$, where a misspecified symmetric model with $\Delta=0$ is substituted into \eqref{eq:bisquare1D} and hence into \eqref{eqn:cov1}. Since the misspecification is in the interaction function, their parameters $A$ and $r$, with $\Delta=0$, were re-estimated by maximum likelihood based on $Z_1$ and $Z_2$. As seen in Fig.~\ref{fig:sim}, right panel, the cokriging predictor $\hat\Yvec_1$ is representative of the true process $\Yvec_1$, even where it is not observed. However, the kriging predictor $\widetilde \Yvec_1$ can only shrink to the mean, $\E\{Y_1(\cdot)\} = 0$, in spatial regions where there are no observations; and $\Yvec_1^\dagger$, which is based on a misspecified symmetric model, is clearly a very poor predictor. Cokriging prediction of the dependent variable $Y_2$ is omitted here for the sake of brevity.


\subsection{Deriving classes of cross-covariance functions from marginal covariance functions}\label{sec:cross-cov}

Our conditional approach may also be used to complement the joint approach to constructing multivariate covariance functions. In particular, \citet{GentonKleiber2015} posed an open problem that seems difficult when using a joint approach: ``[G]iven two marginal covariances, what is the valid class of possible cross-covariances that still results in a nonnegative-definite structure?''. A straightforward answer is available through our conditional approach. The class of cross-covariance functions is given by \eqref{eqn:cov2} for any integrable function $b(\svec,\v)$ such that the function $C_{2\mid 1}(\cdot,\cdot)$ obtained from \eqref{eqn:cov1} is nonnegative-definite. This is potentially a very rich class of cross-covariance functions, and answering the question reduces to verifying which choice of $b(\cdot,\cdot)$ in \eqref{eqn:cov1} yields a nonnegative-definite $C_{2|1}(\cdot,\cdot)$.

For example, consider the stationary case in $D = \mathbb{R}^2$, where we have stationary covariance functions $C_{11}(\h), C_{2\mid 1}(\h)$, and interaction function $b(\s,\v) = b_o(\v - \svec)$. Then from \eqref{eqn:cov1},
\begin{equation*}
C_{2\mid 1}(\h) = C_{22}(\h) - \int_{\mathbb{R}^2}\int_{\mathbb{R}^2}{b_o(\tilde\v)b_o(\tilde\w)C_{11}(\h - \tilde\v + \tilde\w)\,\intd\tilde\v\intd\tilde\w}.
\end{equation*}
\noindent Let $\omegab \in \mathbb{R}^2$ denote spatial frequency, and let $\Gamma_{11}(\omegab), \Gamma_{22}(\omegab)$, and $B_o(\omegab)$ be the Fourier transforms of $C_{11}(\h), C_{22}(\h)$, and $b_o(\h)$, respectively. Then, for $C_{2\mid 1}(\h)$ to be a valid covariance function, it is required that $\Gamma_{22}(\omegab) - B_o(\omegab)B_o(-\omegab)\Gamma_{11}(\omegab)$ be nonnegative and integrable over $\omegab \in \mathbb{R}^2$ \citep{Cressie_1999,Gneiting_2002}. The nonnegativity is trivial if $\Gamma_{11}(\omegab) = 0$, hence consider those $\omegab \in \Omega$ for which 
\begin{equation}\label{eq:B_ineq}
B_o(\omegab)B_o(-\omegab) \le \Gamma_{22}(\omegab)/\Gamma_{11}(\omegab),
\end{equation}
where $\Gamma_{11}(\omegab) > 0$. Recall that $C_{11}(\h)$ and $C_{22}(\h)$ are covariance functions and hence, necessarily, $\Gamma_{11}(\omegab) \ge 0$ and $\Gamma_{22}(\omegab) \ge 0$. Further, $B_o(\omegab)B_o(-\omega)\ge 0$, trivially.

Any $B_o(\cdot)$ that satisfies \eqref{eq:B_ineq} gives the required result, because finiteness follows from $\int\Gamma_{22}(\omegab)\,\intd\omegab < \infty$ being an upperbound on the integral, $\int \Gamma_{22}(\omegab) - B_o(\omegab)B_o(-\omegab)\Gamma_{11}(\omegab)\,\intd\omegab$. Notice that $\Gamma_{11}(\cdot)$ and $\Gamma_{22}(\cdot)$ are Fourier transforms of any pair of stationary covariance functions, and that the squared modulus of $B_o$ has only to stay below the envelope given by the right-hand-side of \eqref{eq:B_ineq}. From our conditional approach, we see that there are many solutions to Genton and Keiber's open problem. Appendix 1 shows how to obtain a class of valid Mat{\'e}rn cross-covariance functions developed by \citet{Gneitingetal2010} that satisfies \eqref{eq:B_ineq}.

\section{Multivariate spatial models through conditioning}\label{sec:4}

\subsection{Definition of cross-covariance functions}

In this section, we extend our conditional approach from the bivariate to the multivariate case. Initially, we work with the variables in their original ordering and subsequently show how directed graphical models introduce parsimony into the conditional approach. Now, $[Y_1(\cdot),\dots,Y_p(\cdot)]$ can be decomposed as
\begin{equation}\label{eq:decomp}
[Y_p(\cdot) \mid  Y_{p-1}(\cdot),Y_{p-2}(\cdot),\dots,Y_1(\cdot)]\times[Y_{p-1}(\cdot) \mid  Y_{p-2}(\cdot),\dots,Y_1(\cdot)]\times\dots \times[Y_1(\cdot)].
\end{equation}
First, we set $\cov\{Y_1(\svec),Y_1(\uvec)\} = C_{11}(\svec,\uvec);~\svec, \uvec \in D$. Analogous to the bivariate case $p=2$, we define the first two conditional moments of $Y_q(\cdot)$, for $q = 1,\dots,p$, as
\begin{align}
\E[Y_q(\svec) \mid  \{Y_r(\cdot) : r = 1,\dots,(q-1)\}] &= \sum_{r = 1}^{q-1} \int_D b_{qr}(\svec,\v)Y_r(\v) \intd \v,\quad \svec \in D, \label{eq:E_multi}\\
\cov[Y_q(\svec), Y_q(\uvec) \mid  \{Y_r(\cdot) : r = 1,\dots,(q-1)\}] &= C_{q \mid  (r < q)}(\svec,\uvec),\quad \svec,\uvec \in D, \label{eq:cov_multi}
\end{align}
where $\{b_{qr}(\cdot,\cdot) : r = 1,\dots,(q-1) ;~q = 2,\dots,p\}$ are integrable functions that give the conditional relationship of the $r$th process on the $q$th process, for $r < q$.

As a result of the decomposition in \eqref{eq:decomp}, we obtain from \eqref{eq:E_multi} and \eqref{eq:cov_multi} the following expression for the marginal covariance functions. For $q = 1,\dots,p$,
\begin{align}
C_{qq}(\svec,\uvec) &\equiv \cov\{Y_q(\svec), Y_q(\uvec)\} \nonumber \\
 & = \sum_{r=1}^{q-1} \sum_{r'=1}^{q-1} \int_D\int_D b_{qr}(\svec,\v)C_{rr'}(\v,\w)b_{qr'}(\uvec,\w)\intd\v\intd\w + C_{q \mid  (r<q)}(\svec,\uvec). \label{eq:Cqq}
\end{align}
Once again, we see that the covariation, here given by~\eqref{eq:Cqq}, is decomposed into an explanatory component and a descriptive component.

For $r=1,\dots,q-1$, the cross-covariance functions are
\begin{align}
C_{rq}(\svec,\uvec) &\equiv \cov\{Y_r(\svec), Y_q(\uvec)\} = \sum_{r'=1}^{q-1} \int_D\int_D b_{qr'}(\uvec,\w)C_{rr'}(\svec,\w)\intd\w, \label{eq:Crq}
\end{align}
and $C_{qr}(\svec,\uvec)\equiv C_{rq}(\uvec,\svec)$. Expressions \eqref{eq:Cqq} and \eqref{eq:Crq} depend on $C_{rr'}$, for $r,r' < q$, which are defined iteratively: Starting with $q=2$, $C_{22}$, $C_{12}$, and $C_{21}$ depend on $C_{11}$ and $C_{2\mid 1}$. The same idea is repeated for $q = 3,\dots,p$.


\subsection{Existence of a $p$-variate process}\label{sec:exist2}

Following the discussion in Section \ref{sec:3-1}, the existence of a $p$-variate Gaussian process with covariance and cross-covariance functions given by~\eqref{eq:Cqq} and~\eqref{eq:Crq} follows by showing that 
\begin{equation}\label{eq:Jp}
\var\left\{\sum_{q=1}^p \sum_{k=1}^{n_q} a_{qk}Y_q(\svec_{qk})\right\} \ge 0,
\end{equation}
for any real numbers $\{a_{qk}: k = 1,\dots,n_q;\,q = 1,\dots,p\}$, any nonnegative integers $\{n_q: q=1,\dots,p\}$ such that $n_1 + \dots + n_p > 0$, and any $\{\svec_{qk} : k = 1,\dots,n_q;\,q = 1,\dots, p\}$. In Appendix 2, we demonstrate that \eqref{eq:Jp} is equal to
\begin{equation}\label{eq:Jp2}
\sum_{m=1}^{n_p}\sum_{m'=1}^{n_p}a_{pm}a_{pm'}C_{p \mid  (q < p)}(\svec_{pm},\svec_{pm'}) + \sum_{q=1}^{p-1}\sum_{r=1}^{p-1}\int_D\int_Da_q(\svec)a_r(\uvec)C_{qr}(\svec,\uvec)\intd \svec \intd \uvec,
\end{equation}
\noindent where 
\begin{equation}\label{eq:a_def2}
a_q(\svec) \equiv \left\{\sum_{k=1}^{n_q}a_{qk}\delta(\svec - \svec_{qk}) + \sum_{m=1}^{n_p}a_{pm}b_{pq}(\svec_{pm},\svec)\right\}.
\end{equation}
The nonnegativity of the first term in \eqref{eq:Jp2} follows by assumption, and the nonnegativity of the second term follows by induction; see Appendix 2.

This result implies that a multivariate spatial Gaussian model constructed using the conditional approach \eqref{eq:E_multi} and \eqref{eq:cov_multi} exists, provided that the univariate covariance functions $C_{11}(\cdot,\cdot)$ and $\{C_{q \mid  (r < q)}(\cdot,\cdot): q = 2,\dots,p\}$ are valid and that the interaction functions $\{b_{qr}(\cdot,\cdot) : r = 1,\dots,q-1;~q = 2,\dots,p\}$ are integrable, which are mild restrictions. Moreover, these functions can be specified completely independently of one another.

\subsection{Joint distributions implied by a network}
\label{sec:spatialnetworks}

Ambient air pollution can cause health problems but not the other way around. Both variables exhibit spatio-temporal variabilities, however data are not available to track all of the parcels of air and individuals interacting in a space-time cube. Integrating these two spatio-temporal processes over time, results in a bivariate spatial process. Is a causal relationship still present? 

Suppose that $a_o(h;\tau)$ is a space-time interaction function, where $h$ and $\tau$ denote spatial and temporal separation respectively. Importantly, assume $a_o(h;\tau)$ is zero for $\tau \le 0$. We consider the mean-zero case and express $Y_2(\svec;t)$ as a causative space-time convolution involving $Y_1(\cdot\,;\cdot)$,
\begin{equation}
\label{eq:Y_2def}
Y_2(\svec ; t) = \int_{-\infty}^\infty\int_D Y_1(\v;t-\tau)a_o(\svec-\v;\tau)\intd\v\intd\tau + \intd V(\svec;t),
\end{equation}
where we let $\intd V$ be a mean-zero, Gaussian, temporally uncorrelated process that satisfies $\var\{\intd V(\svec;t)\}$ $= 4|t|\intd t$ and $\intd t$ is an infinitesimal interval at $t$. At each time point $t$, $\intd V(\svec;t)$ is assumed to be spatially correlated in a manner invariant with $t$. A simple example of such a process is one that is space-time separable, where the temporal component is $2|t|^{1/2} \intd W(t)$ for $W(\cdot)$ a Wiener process. Interchanging the order of integration, the time-integrated process is
\begin{equation*}
\lim_{T \rightarrow \infty}\frac{1}{2T}\int_{-T}^T Y_2(\svec;t)\intd t = \lim_{T \rightarrow \infty}\frac{1}{2T} \int_{-T}^T\int_D  \int_{-\infty}^\infty Y_1(\v; t - \tau)a_o(\svec-\v;\tau)\intd \tau \intd \v\intd t + \xi(s),  
\end{equation*}
where it can be shown that the spatial covariance function of $\xi(s)$ is identical to that of $\intd V(\svec;t)$; see \citet[Section 4.2]{daPrato_2014} for a formal treatment. The inner integrand of the first term on the right-hand side is a convolution that is a function of $t$. Applying Fubini's theorem to the convolution \citep[e.g.,][Chapter 6]{Wheeden_2015}, we obtain
\begin{equation}\label{eq:Y2bar}
\overline Y_2(\svec) = \int_D \overline Y_1(\v) b_o(\svec-\v) \intd \v + \xi(s),
\end{equation}
where $\overline Y_q(\svec) \equiv \lim_{T \rightarrow \infty}(2T)^{-1}\int_{-T}^T Y_q(\svec;t)\intd t$, $b_o(\svec - \v) \equiv \lim_{T \rightarrow \infty}\int_{-T}^T a_o(\svec-\v;t)\intd t$, and where one must ensure that $\int_{-\infty}^\infty a_o(\svec-\v;t)\intd t < \infty$, for all $s,v$. Clearly, the spatio-temporal interaction function $a_o(\cdot;\cdot)$ is not identifiable from $b_o(\cdot)$, but the causative structure in \eqref{eq:Y_2def} implies a causative relationship in the spatial domain $D$, from $\overline{Y}_1$ to $\overline{Y}_2$ through $b_o(\cdot)$. Comparing \eqref{eq:Y2bar} to the bivariate model in Section \ref{sec:2}, we can identify $\cov\{\xi(\svec),\xi(\uvec)\}$ with $C_{2|1}(\svec,\uvec)$.

We continue the discretization by tesselating $D$ into small finite elements. Then~\eqref{eq:Y2bar} can be written as $Y_2=BY_1+\xi$, where the elements of the matrix $B$ are defined by discretizing the interaction function. This bivariate model can be represented as a simple directed acyclic graph with $Y_1$ as the parent node and $Y_2$ as the child node. It is straightforward to see that assumptions similar to~\eqref{eq:Y_2def} about the spatio-temporal dependence for $p~(\geq2)$ variables, will engender a directed acyclic graph. This has become an important approach used in multivariate statistical modelling (e.g.,~\citealp{Cox_1996}), and our research in this paper shows how it generalizes to multivariate spatial statistical modelling. The directed acyclic graph structure is equivalent to a partial order on the nodes \citep[e.g.,][]{Cressie_1998}. Then~\citep[p. 362]{Bishop_2006}:
\begin{equation}
[Y_1,\ldots,Y_p]=\prod_{q\in\overline{R}}[Y_q|Y_{pa(q)}][Y_R],\label{eq:Y_qdef}
\end{equation}
where $Y_R$ is set of spatial processes whose indices are given by all the root nodes, $\overline{R}$ are all the nodes with parents, and $pa(q)$ are all the parent nodes that have a directed edge to node $q$.

When there is causative structure between the $p$ variables, expressed through a directed acyclic graph,~\eqref{eq:Y_qdef} shows that the $p!$ possible multivariate models reduces to just one. The modeller then needs to specify and fit the interaction functions, $\{b_{q,pa(q)}(\cdot,\cdot):q\in\overline{R}\}$, and the multivariate marginal distribution $[Y_R]$. The special case of a rooted tree, common in multiresolutional spatial models, leaves just one spatial process, say $[Y_1]$, to model marginally (e.g.,~\citealp{verHoef_1998}; \citealp{Huangetal2002}).  
If there are feedback loops in the space-time cube considered earlier, temporal aggregation will result in undirected edges between the relevant variables.  For these edges, a choice of direction that results in a directed acyclic graph results in a multivariate model. The fewer edges there are that are undirected, the smaller the number of possible multivariate models to fit via the conditional approach. Of course, it is possible to combine nodes of the network until all remaining edges are directed. In that case, $[Y_{Q|pa(Q)}]$ is a $|Q|$-variate conditional model, where $Q$ is the combined node consisting of $|Q|$ spatial variables.

An undirected edge may be due to directed edges from a missing node in the network; for example, exposure to cigarette smoke was a variable missing from the two-node network of~\cite{Jin_2005}, where lung cancer was modelled conditional on esophagus cancer. Without the presence of the third node, a full bivariate modelling approach may seem more appropriate than a conditional approach. Alternatively, an edge may be undirected because the causative mechanism is not yet well understood, and the conditional approach will shed light on this. In this case, both directions can be tried, and a model-selection criterion, such as cross-validation, the Akaike information criterion, or the deviance information criterion, would indicate the appropriate direction of the edge. In Section~\ref{sec:estimation}, we illustrate a case where the directed edge from the temperature variable to the pressure variable is unequivocal. Model selection in this framework amounts to establishing the causative links in the network (e.g.,~\citealp{Lauritzen_1996,Kolaczyk_2009}).


Finally, there are other, more direct ways that could guide the choice of edges in the network of spatial variables. Generally speaking, although not necessarily, $Y_2(\cdot)$ will be a smoother process than $Y_1(\cdot)$, due to the integral in~\eqref{eqn:cov1}. Hence, a Mat\'{e}rn model could be fitted to each individual spatial process and an ordering of the fitted Mat\'{e}rn smoothness parameters could be used to indicate the directed edges. A similar problem involving choice of edges was faced by time-series analysts, where stationarity was assumed and the dependence was captured through a spectral and cross-spectral representation of the process' covariance and cross-covariance functions.~\cite{Dahlaus_2000}  developed a series of hypothesis tests in spectral space to determine undirected edges in a network of temporal processes.

\section{Analyzing a temperature-pressure dataset}\label{sec:5}

\subsection{The data}

We demonstrate the flexibility of the conditional approach on a temperature-pressure dataset used in \cite{Gneitingetal2010} and \cite{Apanasovich_2012}. The data, which are available with the \texttt{R} package \texttt{RandomFields} \citep{Schlachter_2015}, are on the error fields, namely the difference between temperature and pressure two-day forecasts and the respective observations from monitoring stations in the Pacific Northwest of North America on December 18, 2003 at 4 p.m. Since the observations are colocated, $m_1 = m_2 \equiv m = 157$, and $D_1^O = D_2^O \equiv D^O$. Both pressure and temperature forecasts are spatially smooth, although observations of temperature tend to be more variable than those of pressure. The smoothing action of the interaction function in \eqref{eqn:cov1} can capture this, which implies that we should condition on the temperature field. See below, where we diagnosed the suitability of this choice by swapping the roles of temperature and pressure in~\eqref{eqn:E-and-cov}.


\subsection{The processes and their bivariate models}

Here we discuss the spatial processes involved with temperature; a discussion of pressure follows likewise. There is a latent temperature process $T(\cdot)$ for which we have observations, $O_1(s_i)=T(s_i)+e_{1,O}(s_i)$, and forecasts, $F_1(s_i)=T(s_i)+e_{1,F}(s_i)~(i=1,\dots,157)$, where $e_{1,O}(\s_i)$ and $e_{1,F}(s_i)$ are the observation and forecast errors, respectively. Then, the data are $Z_1(s_i)=Y_1(s_i)=F_1(s_i)-O_1(s_i)=e_{1,F}(s_i)-e_{1,O}(s_i)~(i =1,\dots,157)$. Notice that the process $Y_1(\cdot)$ itself is defined in terms of observations, so we analyze the problem assuming that the temperature data $Z_1$ and the process $Y_1$ at their respective locations are the same. We do likewise for pressure, resulting in data $\Zvec_2$ and the process $Y_2$ the same as $Z_2$.

For both variables, the data are incomplete and hence cokriging is needed to map the respective fields. Specifically, for any $s_0\in D$, our goal is to use cokriging to predict $Y_1(s_0)$ and $Y_2(s_0)$ and to compute their prediction standard errors. In what follows, a number of bivariate spatial models based on the conditional approach are fitted and their performances compared to Mat\'{e}rn-type models fitted by~\cite{Gneitingetal2010}.


In the conditional approach given by~\eqref{eqn:E-and-cov}--\eqref{eqn:C11}, we need to specify the univariate covariance functions, $C_{11}(\svec,\uvec)$ and $C_{2\mid 1}(\svec,\uvec)$, and the integrable interaction function $b(\svec,\v)$. We let the covariance functions be isotropic Mat{\'e}rn covariance functions given by \eqref{eq:Matern1} and \eqref{eq:Matern2}. Further, we let $b(\svec,\v)$ be a function of displacement, $\h \equiv \v - \svec$, so that $b_o(\h) \equiv b(\svec,\v)$. The four different models fitted are written as:
\begin{equation*}
\begin{array}{ll}
\textrm{Model 1 (independent processes):} &b_o(\h) \equiv 0, \\
\textrm{Model 2 (pointwise dependence):} &b_o(\h) \equiv A\delta(\h), \\
\textrm{Model 3 (diffused dependence):} &b_o(\h) \equiv \left\{\begin{array}{ll} A\{1 - (\|\h\|/r)^2\}^2, & \qquad \| \h\| \le r, \\ 0, & \qquad \textrm{otherwise,} \end{array} \right.  \\ 
\textrm{Model 4 (asymmetric dependence):} &b_o(\h) \equiv \left\{\begin{array}{ll} A\{1 - (\|\h - \Deltab\|/r)^2\}^2, & \| \h - \Deltab\| \le r, \\ 0, & \textrm{otherwise,} \end{array} \right.
\end{array}
\end{equation*}
where $b_o(\cdot)$ in Models 3 and 4 is a bisquare and shifted bisquare function, respectively. The introduction of the asymmetric parameter $\Deltab=(\Delta_1, \Delta_2)^\T$ in Model 4 is analogous to applying the shifting method of \citet{VerHoefCressie1993}, \citet{ChristensenAmemiya2001}, and \citet{Li_2011} to the diffused symmetric dependence in Model 3. We explored whether asymmetry might be present in the data by first interpolating the temperature and pressure error fields onto a regular grid, and then plotting the correlation between the two gridded fields as a function of the displacement vector $h$ of the temperature field. The contour plot showed a clear negative dip in the bottom-right quadrant, indicating asymmetry. In contrast, the bivariate spatial models fitted to these data by \citet{Gneitingetal2010} and \citet{Apanasovich_2012} are symmetric.

\begin{figure}[!t]
\includegraphics[width=\textwidth]{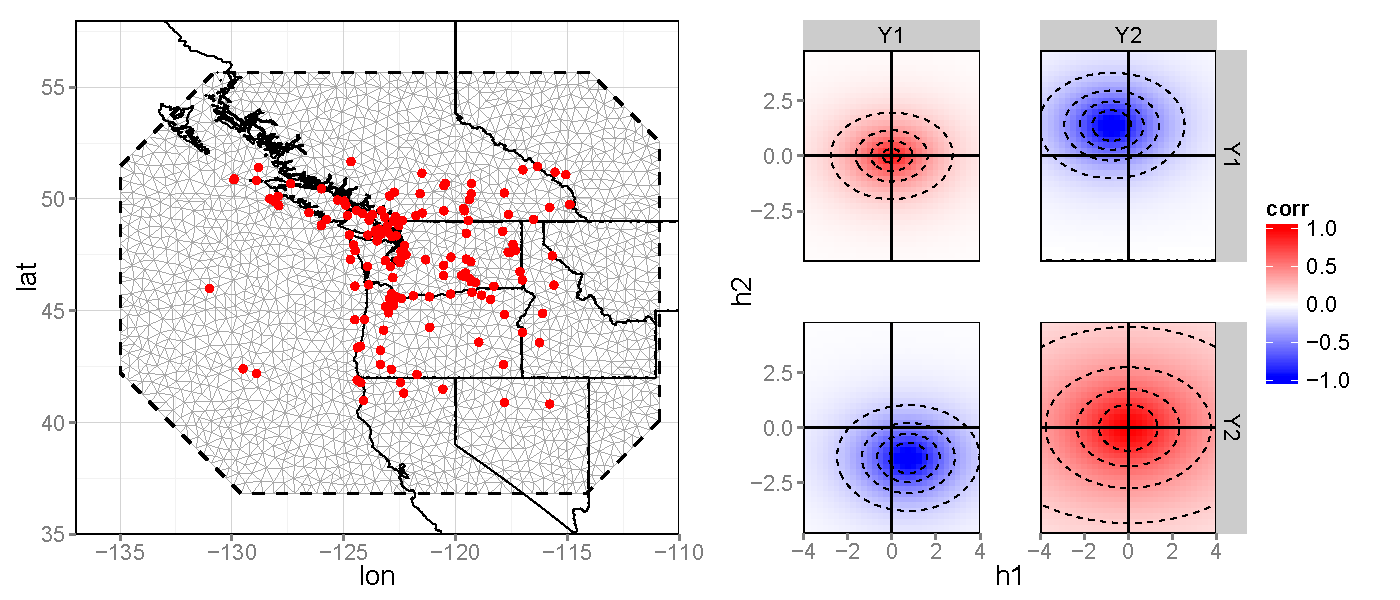}
 	\caption{Spatial domain and correlation functions. Left panel: State boundaries and province boundaries of a region of the USA and Canada (dark solid lines), with the domain of interest enclosed by a bounding polygon (dashed line). The irregular triangular grid used for discretizing $D$ (light solid lines) and the observation locations given by $D^O$ (dots) are also shown. The discretized spatial domain $D^L$ consists of the vertices of the triangular grid. Right panel: The correlation and cross-correlation functions estimated using Model 4, depicted as a function of displacement $h$, in degrees longitude/latitude, at the location $\svec = (-123^\circ,45^\circ)$. Contour lines of correlation are in intervals of $0.2$.} 
\label{fig:mesh}
 \end{figure}

We discretized both $Y_1(\cdot)$ and $Y_2(\cdot)$ onto a triangulated grid using the mesher available with the \texttt{R} package \texttt{INLA} available from www.r-inla.org. The resulting irregular spatial lattice had $n_1= n_2 \equiv n = 2063$ vertices each. Here, these $n$ vertices define $D^L$; see Fig.~\ref{fig:mesh}, left panel.
Under the chosen triangulation, the integral in \eqref{eqn:E-and-cov} is approximated as $\E\{Y_2(\svec_l) \mid  Y_1(\cdot)\} \simeq \sum_{k=1}^{n} \eta_k b(\svec_l,\v_k)Y_1(\v_k)$, where in this case $\{\eta_k: k = 1,\dots,n\}$ are the areas of the small Voronoi polygons constructed from the triangulated grid \citep[e.g.,][]{Lee_1980}. In order to ensure nonnegative-definiteness of $C_{11}$ and $C_{2|1}$, we follow~\citet{Gneitingetal2010} and use chordal distances to establish the covariances between two points on the sphere. This embeds Earth's surface into $\mathbb{R}^3$, where univariate covariance functions are readily available. The interaction function has no such constraint, so we capture asymmetry in the interaction function $b_o(\cdot)$ directly in the longitude-latitude space and carry out the numerical integration there.

\subsection{Estimation and prediction}\label{sec:estimation}

From \eqref{eqn:cov-matrix}, the covariance matrix of the bivariate spatial process is
\begin{equation}
\cov((Y_1^{\T},Y_2^{\T})^\T) = \begin{bmatrix}\bSigma_{11} + \tau^2_1I_{m}& \bSigma_{11}\bB^\T \\ \bB \bSigma_{11} & \bSigma_{2\mid 1}+\bB\bSigma_{11}\bB^\T + \tau^2_2I_{m}  \end{bmatrix}, \label{eq:cov_Ydash}
\end{equation} 
which is a $4126\times 4126$ matrix. The terms $\tau^2_1\Imat_m$ and $\tau^2_2\Imat_m$ are due to micro-scale effects, which we add to make our model comparable with that of \cite{Gneitingetal2010}. Maximum likelihood estimation took on the order of 1 minute for Models 1 and 2, and on the order of 1 hour for Models 3 and 4. Computational requirements when numerical integration is required can be reduced by using a covariance function $C_{11}$ that can be evaluated rapidly on a fine grid, or that has compact support \citep{Furrer_2012}. 

The maximum likelihood estimates of the parameters for the four different interaction functions are given in Table \ref{tab:results}. Notice that some of the estimates change considerably between model specifications. For example, the scale parameter $\sigma_{2\mid 1}$ decreases from $275.34$ in the independent model to $199.86$ in the asymmetric-dependence model, which illustrates how some of the variability in the pressure error field is accounted for by conditioning on the temperature error field. The estimate of the interaction parameter $A$ is also seen to become steadily more negative from Model 2 to Model 4, implying that the interaction function is most influential when it is allowed to have both a scale and an asymmetry term. 

\begin{table}[t!]
\caption{Parameter estimates for Models 1--4. Blank entries indicate that the parameter is not present in the model.}
\vspace{0.1in}
\resizebox{\columnwidth}{!}{%
\begin{tabular}{rrrrrrrrrrrrr}
  & $\tau_1$ & $\tau_2$ & $\sigma_{11}$ & $\sigma_{2|1}$ & $\kappa_{11}$ & $\kappa_{2|1}$ & $\nu_{11}$ & $\nu_{2|1}$ & $A$ & $r$ & $\Delta_1$ & $\Delta_2$\vspace{0.07in}\\ 
Model 1 & 0.00 & 68.47 & 2.60 & 275.34 & 0.011 & 0.010 & 0.60 & 1.56 &  &  &  &  \\ 
  Model 2 & 0.00 & 67.78 & 2.60 & 242.04 & 0.011 & 0.011 & 0.60 & 1.58 & $-$14.30 &  &  &  \\ 
  Model 3 & 0.00 & 70.16 & 2.68 & 243.77 & 0.011 & 0.010 & 0.61 & 1.84 & $-$40.83 & 1.46 &  &  \\ 
  Model 4 & 0.01 & 69.79 & 3.02 & 199.86 & 0.007 & 0.004 & 0.56 & 1.24 & $-$65.58 & 1.18 & 0.76 & $-$1.42 \\ 
  \end{tabular}}
\label{tab:results}
\end{table}

\begin{table}[t!]
\def~{\hphantom{0}}
\caption{Log-likelihood (Log-lik.), Akaike information criterion (AIC) for Models 1--4, the parsimonious Mat{\'e}rn model,  the shifted parsimonious Mat{\'e}rn model, and the full Mat{\'e}rn model}
\vspace{0.1in}
\begin{tabular}{rrrr}
  & No. of parameters & Log-lik. & AIC \vspace{0.07in}\\ 
  Model 1 & 8 &   $-$1276.77 & 2569.54\\ 
  Model 2 & 9 &  $-$1269.92 & 2557.84 \\ 
  Model 3 & 10 &  $-$1264.90 & 2549.80 \\ 
  Model 4 & 12 &  $-$1258.21 & 2540.43\vspace{0.07in}\\
  Parsimonious Mat{\'e}rn & 8 & $-$1265.76 & 2547.52 \\
  Shifted parsimonious Mat{\'e}rn & 10 & $-$1260.87 & 2541.75 \\
  Full Mat{\'e}rn & 11 & $-$1265.53 & 2553.06
  \end{tabular}

\label{tab:results2}
\end{table}

Since out-of-sample spatial prediction is a principal use of multivariate spatial models, we used the Akaike information criterion and cross-validation to assess model performance \citep{Stone_1977}. As seen in Table \ref{tab:results2}, the Akaike information criterion decreases steadily from Model 1 with 8 parameters ($2569.54$) to the lowest at Model 4 with 12 parameters ($2540.43$). The symmetric Mat{\'e}rn models of \citet[][Table 3]{Gneitingetal2010} 
performed worse than Model 4, while the parsimonious Mat{\'e}rn model gave a similar Akaike information criterion to Model 3. The shifted parsimonious Mat{\'e}rn model, constructed by applying the method of \citet{Li_2011} to the parsimonious Mat{\'e}rn model, gave a similar Akaike information criterion to Model 4. These results were expected due to the similarity between Model 3 and the parsimonious Mat{\'e}rn model, which is found in Appendix 1, and due to the analogy between the approach of \citet{Li_2011} and our inclusion of $\Delta$ in Model 4. Overall, these results suggest that allowing for asymmetries in the model is more important in this problem than incorporating smoothness and/or scale parameters in the cross-dependencies. Correlation and cross-correlation functions estimated from Model 4 are shown in Figure~\ref{fig:mesh}, right panel.

For our cross-validation analysis we left out a single location and found the predictive distribution of both fields at the left-out location using parameters estimated from all the data. In Table \ref{tab:LOO}, which is found in Appendix 3, we list the maximum absolute error, the root-mean-squared prediction error, and the mean continuous-ranked probability score from our cross-validation study. As well as giving results for Models 1--4, we also include those obtained using the parsimonious Mat{\'e}rn and the full bivariate Mat{\'e}rn models with the \texttt{RandomFields} package \citep{Schlachter_2015} and those from the parsimonious Mat{\'e}rn model. Within the models we propose, Model 4 and the shifted parsimonious Mat{\'e}rn model outperformed the others on nearly all cross-validation diagnostics for both the pressure and temperature error fields. When compared to the symmetric parsimonious and full bivariate Mat{\'e}rn models, the asymmetric models offer considerable improvement in the prediction performance of both error fields.

\begin{figure}[!t]
\includegraphics[width=\textwidth]{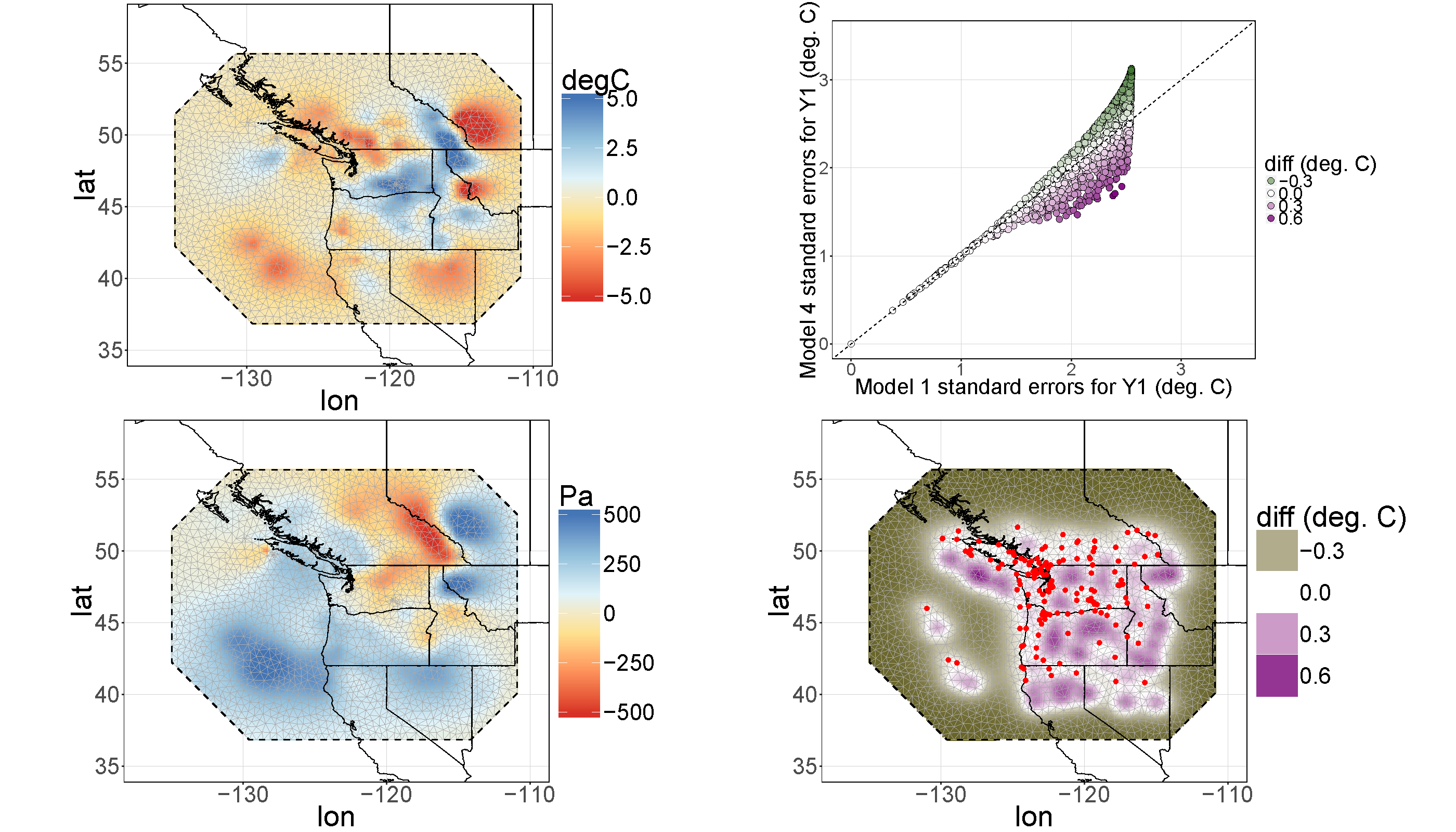}
 	\caption{Cokriging predictions in the discretized spatial domain. Left panels: The cokriged surface using maximum likelihood estimates for the parameters with Model 4 for the temperature and pressure error fields. Top-right panel: A scatter plot of the cokriging prediction standard errors of $Y_1$ obtained with Model 4 against those obtained with Model 1 at each of the mesh vertices. The colour illustrates the difference between the two, with green denoting the higher standard error of Model 4 and purple denoting the higher standard error of Model 1. Bottom-right panel: A spatial plot of the difference in the prediction standard errors of $Y_1$ obtained with Model 4 and Model 1, with green denoting a higher standard error of Model 4 and purple denoting a higher standard error of Model 1.} \label{fig:kernel}
 \end{figure}

In Fig. \ref{fig:kernel}, left panels, we show the cokriged temperature and pressure error fields under Model 4 using the entire dataset that includes both $Z_1$ and $Z_2$. Notice how the temperature error field is considerably rougher than the pressure error field. In Fig. \ref{fig:kernel}, right panels, we illustrate for temperature the difference between the cokriging standard errors based on Model 1 and those based on Model 4. The spatial pattern of the standard errors is a clear consequence of the asymmetric covariance function: comparing Models 1 and 4, we see that Model 4 tends to have lower standard errors in regions that are south-east of the observation locations, which is due to Model 4's asymmetry.

Finally, we re-did all the experiments for the same models described above, but now with $Y_1$ as pressure error and $Y_2$ as temperature error. With this reversed conditioning, the Akaike information criteria for Models 1--3 did not change substantially, however that for Model 4 worsened from $2540.43$ to $2560.97$. Further, the analogous leave-one-out cross-validation diagnostics showed that the reversed modelling of temperature error given pressure error in Model 4 resulted in worse predictive performance in the respective entries of Table \ref{tab:LOO}, with regard to both temperature and pressure. Clearly, the direction of dependence plays a central role here in model performance and our conditional approach has allowed us to propose a preferred direction. We discuss this in greater detail in Section~\ref{sec:spatialnetworks}.

\section{Discussion}\label{sec:6}

The conditional approach can be modified easily for different spatial domains. Consider $\{Y_1(\svec): \svec \in D_1\}$ and $\{Y_2(\svec): \svec \in D_2\}$, for $D_1, D_2 \subset \mathbb{R}^d$; then \eqref{eqn:E-and-cov} becomes,
\begin{equation*}
\E(Y_2(\svec) \mid  Y_1(\cdot)) = \int_{D_1}b(\svec,\v)Y_1(\v)\intd\v;\quad\svec \in D_2.
\end{equation*}
\noindent For example, \citet[][p.~287]{CressieWikle2011} illustrate bivariate spatial dependence between Mallard breeding bird pairs in the Prairie Pothole region of North America and the El Ni\~{n}o phenomenon in the tropical Pacific Ocean, for which the conditional approach could be used. 

In Section \ref{sec:5}, we estimated the parameters appearing both in $C_{11}(\h)$, $C_{2\mid 1}(\h)$, and in the interaction function $b(s,v)$. In some cases, $b(s,v)$ can be given by the underlying science. One such case is atmospheric trace-gas inversion \citep{Zammit_2016}, in which a non-Gaussian flux field $Y_1$ is estimated from the mole-fraction field $Y_2$, observed at isolated locations. The interaction function $b(s,v)$ was obtained directly from a transport model driven by weather forecasts and hence was assumed known \citep[e.g.,][]{Ganesan_2014}. 

Even if the parameters are known or estimated off-line, spatial or spatio-temporal inference with multivariate models can remain computationally challenging. When treating all variates simultaneously in joint form, sparse formulations and sparse linear-algebraic methods can greatly facilitate the computation \citep[e.g.,][]{Zammit_2015a}. Sparseness is guided by the graphical representations, which are discussed in Section~\ref{sec:spatialnetworks}. By constructing multivariate spatial models through conditioning, the accompanying graphical representations allow exact inference through sequential algorithms. Markov chains of spatial processes, such as autoregressive spatio-temporal processes, can be tackled with the iterative Rauch--Tung--Striebel smoother \citep[e.g.,][]{Rauch_1965}. For more general constructions, such as trees or polytrees, the sum-product or peeling algorithm may be used for exact inference. When likelihoods associated with some or all of the processes in $\{Y_q : q = 1,\dots,p\}$  are intractable, approximate message passing may be used to keep the computations tractable \citep[e.g.,][]{Heskes_2002}, such as when the data model for $Z_q(\cdot)$ is a spatial Poisson point process and $Y_q(\cdot)$ is the log-intensity of the process.

Reproducible code and data are available from https://github.com/andrewzm/bicon.

\section*{Acknowledgment}

We would like to thank Chris Wikle, Sudipto Banerjee, and Kai Du for discussions on the conditional approach to modelling multivariate spatial dependence, as well as the the referees and editor for their constructive comments. This research was partially supported by the U.S. National Science Foundation (NSF) and the U.S. Census Bureau 
through the NSF-Census Research Network (NCRN) program; and it was partially supported by a 2015--2017 Australian Research Council Discovery Project.

\appendix

\section*{Appendix 1}
\subsection*{A class of Mat{\'e}rn cross-covariance functions consistent with marginal Mat{\'e}rn covariance functions}

Let $C_{11}(\h), C_{22}(\h)$, and $b_o(\h)$ be isotropic Mat{\'e}rn covariance functions on $\mathbb{R}^2$ and, for simplicity, assume that they all have the same scale $\kappa$. Then, using obvious notation, their Fourier transforms are 
\begin{align*}
B_o(\omegab) &= \sigma^2_b\frac{\Gamma(\nu_b + 1) \kappa^{2\nu_b}}{\pi\Gamma(\nu_b)}(\kappa^2 + \|\omegab\|^2)^{-\nu_b - 1}, \quad \omega \in \mathbb{R}^2, \\
\Gamma_{ii}(\omegab) &= \sigma^2_{ii}\frac{\Gamma(\nu_{ii} + 1) \kappa^{2\nu_{ii}}}{\pi\Gamma(\nu_{ii})}(\kappa^2 + \|\omegab\|^2)^{-\nu_{ii} - 1}, \quad \omega \in \mathbb{R}^2,~i = 1,2.
\end{align*}
\noindent For $C_{21}(\cdot)$ and $C_{12}(\cdot)$ to be valid cross-covariance functions, it is required that $$\Gamma_{22}(\omegab) - B_o(\omegab)B_o(-\omegab)\Gamma_{11}(\omegab) \ge 0,$$ and hence that
\begin{equation}\label{eq:sigmab_ineq}
\sigma_b^4 \le \frac{\pi^2\sigma^2 _{22}}{\sigma^2_{11}}\frac{1}{\nu_b^2\kappa^{4\nu_b}}\frac{\nu_{22}\kappa^{2\nu_{22}}}{\nu_{11}\kappa^{2\nu_{11}}}(\kappa^2 + \|\omegab\|^2)^{2 + 2\nu_b + \nu_{11} - \nu_{22}}.
\end{equation}
\noindent It can be easily shown that the inequalities,
\begin{align}
\nu_b &\ge (\nu_{22} - \nu_{11} -2)/2, \label{eq:cond2}  \\
\sigma_b^2 &\le 2\pi\frac{\sigma_{22}}{\sigma_{11}}\frac{1}{\nu_{22} - \nu_{11} - 2}\frac{\kappa^{\nu_{22}}}{\kappa^{\nu_{11}}\kappa^{2\nu_{b}}}\left(\frac{\nu_{22}}{\nu_{11}}\right)^{\frac{1}{2}}, \label{eq:sigma2b}
\end{align}
\noindent are sufficient for \eqref{eq:sigmab_ineq} to hold. Then, from \eqref{eqn:cov2}, $C_{12}(\h)$ is also a Mat{\'e}rn covariance function with variance 
\begin{equation}\label{eq:margvar}
\sigma^2_{12} = \frac{1}{\pi\kappa^{2}}\frac{\nu_b\nu_{11}}{\nu_b + \nu_{11} + 1}\sigma_b^2\sigma_{11}^2,
\end{equation}
 and smoothness $\nu_{12} \equiv \nu_b + \nu_{11} + 1$. Hence, from \eqref{eq:cond2}, $\nu_{12} \ge (\nu_{11} + \nu_{22})/2.$


Now consider the bound on the smoothness, $\nu_{12} = (\nu_{11} + \nu_{22})/2,$ which is obtained from the bound, $\nu_b = (\nu_{22} - \nu_{11} - 2) / 2$, in \eqref{eq:cond2}. An inequality for the variance $\sigma^2_{12}$ is then obtained by substituting this value of $\nu_b$ and the inequality \eqref{eq:sigma2b} into \eqref{eq:margvar}: $\sigma^2_{12} \le 2\sigma_{11}\sigma_{22}(\nu_{11}\nu_{22})^{1/2}(\nu_{11} + \nu_{22})^{-1}$. The conditions on $\nu_{12}$ and $\sigma^2_{12}$ are those that \citet{Gneitingetal2010} impose in order to construct parsimonious bivariate Mat{\'e}rn models. Clearly, these are more restrictive than our conditions \eqref{eq:cond2} and \eqref{eq:sigma2b}.

 
 Generalizing these ideas to arbitrary scale parameters $\kappa_{11}, \kappa_{22}, \kappa_b$, as in \citet{Gneitingetal2010} could be done, but it is more fruitful to give up the assumption that the interaction function is a Mat\'{e}rn symmetric nonnegative-definite covariance function; recall that it only needs to be integrable.



\section*{Appendix 2}
\subsection*{Proof of existence  of the multivariate process}

Here, we prove by induction that \eqref{eq:Jp} holds for for any real numbers $\{a_{qk}: k = 1,\dots,n_q; q = 1,\dots,p\}$, any nonnegative integers $\{n_q: q=1,\dots,p\}$ such that $n_1 + \dots + n_p > 0$, and any $\{\svec_{qk} : k = 1,\dots,n_q;~q = 1,\dots, p\}$. We have already shown, through \eqref{eqn:left-hand-of-var}, that there exists a bivariate stochastic process, and hence the variance of any linear combination of the two processes is nonnegative. Now, assume that $\{Y_1(\cdot),\dots,Y_{p-1}(\cdot)\}^{\T}$ is a well defined $(p-1)$-variate stochastic process. We re-write \eqref{eq:Jp} as:
\begin{equation*}
\var\left\{\sum_{q=1}^{p-1} \sum_{k=1}^{n_q} a_{qk}Y_q(\svec_{qk}) + \sum_{m=1}^{n_p}a_{pm}Y_p(\svec_{pm})\right\}.
\end{equation*}

Then, following the definitions for the marginal and cross-covariances in \eqref{eq:Cqq} and \eqref{eq:Crq} and using standard identities, we obtain the following expression for \eqref{eq:Jp}:

\begin{align}
\quad&  \sum_{m=1}^{n_p}\sum_{m'=1}^{n_p}a_{pm}a_{pm'}C_{p \mid  (q < p)}(\svec_{pm},\svec_{pm'})  \nonumber \\
  &~~+\quad \sum_{q=1}^{p-1}\sum_{r=1}^{p-1}\sum_{m=1}^{n_p}\sum_{m'=1}^{n_p} a_{pm}a_{pm'}\int_D\int_Db_{pq}(\svec_{pm},\v) C_{qr}(\v,\w)b_{pr}(\svec_{pm'},\w)\intd\v\intd\w  \nonumber \\
  &~~+\quad \sum_{q=1}^{p-1}\sum_{r=1}^{p-1}\sum_{k=1}^{n_q}\sum_{m'=1}^{n_p} a_{qk}a_{pm'}\int_D b_{pr}(\svec_{pm'},\w)C_{qr}(\svec_{qk},\w)\intd \w  \nonumber \\
  &~~+\quad \sum_{q=1}^{p-1}\sum_{r=1}^{p-1}\sum_{k'=1}^{n_q}\sum_{m=1}^{n_p} a_{qk'}a_{pm}\int_D b_{pq}(\svec_{pm},\v)C_{qr}(\v,\svec_{rk'})\intd \v  \nonumber \\
  &~~+\quad \sum_{q=1}^{p-1}\sum_{r=1}^{p-1}\sum_{k=1}^{n_q}\sum_{k'=1}^{n_r} a_{qk}a_{rk'}C_{qr}(\svec_{qk},\svec_{rk'}), \nonumber 
\end{align}
which can be simplified to
\begin{align}
\quad& \sum_{m=1}^{n_p}\sum_{m'=1}^{n_p}a_{pm}a_{pm'}C_{p \mid  (q < p)}(\svec_{pm},\svec_{pm'})  \nonumber \\  
  &~~+\quad \sum_{q=1}^{p-1}\sum_{r=1}^{p-1}\int_D\int_D \left\{\sum_{k=1}^{n_q}a_{qk}\delta(\svec - \svec_{qk}) + \sum_{m=1}^{n_p}a_{pm}b_{pq}(\svec_{pm},\svec)\right\} \label{eq:temp11}\\
  &~~\quad~~~~~~~~~~~~~~~~~~\times \left\{\sum_{k'=1}^{n_q}a_{rk'}\delta(\uvec - \svec_{rk'}) + \sum_{m'=1}^{n_p}a_{pm'}b_{pr}(\svec_{pm'},\uvec)\right\}C_{qr}(\svec,\uvec)\intd \svec \intd \uvec. \nonumber
\end{align}
Expression \eqref{eq:temp11} can be further reduced to
\begin{align}
  & \sum_{m=1}^{n_p}\sum_{m'=1}^{n_p}a_{pm}a_{pm'}C_{p \mid  (q < p)}(\svec_{pm},\svec_{pm'}) + \sum_{q=1}^{p-1}\sum_{r=1}^{p-1}\int_D\int_Da_q(\svec)a_r(\uvec)C_{qr}(\svec,\uvec)\intd \svec \intd \uvec, \label{eq:App2_Jp}
\end{align}
and this is~\eqref{eq:Jp2}. The first term in \eqref{eq:App2_Jp} is nonnegative by assumption, while the second term is nonnegative since $\{Y_1(\cdot),\dots,Y_{p-1}(\cdot)\}^{\T}$ is a well-defined $(p-1)$-variate process.

\clearpage

\section*{Appendix 3}
\subsection*{Leave-one-out cross-validation diagnostics}
\begin{table}[hb!]
\def~{\hphantom{0}}
\caption{Leave-one-out cross-validation prediction diagnostics: mean absolute error (MAE), root-mean-squared prediction error (RMSPE), and mean continuous-ranked probability score (MCRPS).}
\vspace{0.1in}
\begin{tabular}{llrrr}
\smallskip
 Process & Model & MAE & RMSPE & MCRPS  \\
\multirow{7}{*}{Pressure (Pa)}  & Model 1 & 69.56 & 123.36 & 55.33 \\ 
   & Model 2 & 70.19 & 124.4 & 55.64 \\ 
   & Model 3 & 70.32 & 123.0 & 55.19 \\ 
   & Model 4 & 66.07 & 114.7 & 51.73 \\ 
   & Parsimonious Mat{\'e}rn & 70.15 & 123.0 & 55.35 \\ 
   & Shifted parsimonious Mat{\'e}rn & 67.01 & 115.0 & 52.48\\ 
\smallskip
   & Full Mat{\'e}rn & 66.19 & 122.8 & 55.23 \\ 
  \multirow{7}{*}{Temperature ($^o$C)}   & Model 1 & 1.14 & 1.63 & 0.81 \\ 
   & Model 2 & 1.14 & 1.63 & 0.81 \\ 
   & Model 3 & 1.10 & 1.53 & 0.78 \\ 
   & Model 4 & 1.08 & 1.47 & 0.77 \\ 
   & Parsimonious Mat{\'e}rn & 1.11 & 1.56 & 0.79 \\ 
   & Shifted parsimonious Mat{\'e}rn & 1.09 & 1.48 & 0.77 \\ 
   & Full Mat{\'e}rn & 1.11 & 1.58 & 0.79 \\    
  \end{tabular}
\label{tab:LOO}
\end{table}

\bibliographystyle{biometrika}
\bibliography{../vignettes/Bibliography}

\end{document}